# High-precision terahertz frequency modulated continuous wave imaging method using continuous wavelet transform

## Yu Zhou

School of optical and electronic information, Wuhan National Laboratory for Optoelectronics, Huazhong University of Science and Technology, Wuhan, China, 430074

**Abstract**. Inspired by the extensive application of terahertz imaging technologies in the field of aerospace, we exploit a terahertz frequency modulated continuous wave imaging method with continuous wavelet transform algorithm to detect a multilayer heat shield made of special materials. This method uses the frequency modulation continuous wave system to catch the reflected terahertz signal and then processing the image data by the continuous wavelet transform with different basis functions. By calculating the sizes of the defects area in the final images and then comparing the results with real samples, a novel and practical high-precision terahertz imaging method are demonstrated. Our method can be an effective tool for the terahertz nondestructive testing of composites, drugs and some cultural heritages.

## 1    Introduction

Terahertz (THz) imaging as a kind of non-destructive, non-contact, and non-ionization imaging technology, has shown a great potential in the detection of some special materials.[1] Since THz could penetrate many nonmetallic materials that are opaque in the scope of infrared and visible, and it's not as radioactive as X-rays or attenuating fast like ultrasound, which is used extensively to detect the defects of air hole inside the heat shield in the field of aerospace. [2-4] The heat shield is made of the composite material that traditional detection method such as ultrasonic and infrared can't penetrate while THz imaging could provide the hidden information of such materials. In addition, although X-ray has strong penetration, it can't distinguish chemical information of the composite materials. With the structure of material becoming more and more complex, the need for high precision imaging in various noisy environments has increased. Therefore, in order to meet the requirement of application, our article aims to apply a new frequency modulated continuous wave (FMCW) THz imaging method combined with the continuous wavelet transform (CWT) to achieve high-precision imaging and quantization of the internal defects inside the heat shield. In the aspect of imaging, the FMCW interferometry, which was used in electronic radar half a century ago, has been successfully introduced into optics in recent years. But until now, it is still not widely used in the field of THz imaging. Compared to the traditional time-domain spectroscopy (TDS) system used for THz imaging, the system using FMCW reduced the requirement of response to the signal reception due to the signal measured was the beat frequency with a lower frequency,[5-9] and it indeed enhanced the measurement accuracy. Unlike single-frequency detection, FMCW can obtain the three dimensional information of the object rather than a single plane. In the aspect of algorithm, the CWT is often used to decompose the function of continuous time into wavelet. Compared with the Fourier transform, the CWT can handle the time-frequency signal very well.[13-14] Therefore, it is very efficient in complex information processing such as edge and corner detection. The CWT is also very resistant to noise in the signal. In our study, we first manufacture a heat shield containing artificial air defects with various sizes. Then, the THz FMCW system that we proposed is used to perform a preliminary scan of the heat shield. After that, The CWT with three different basis functions are applied to deal with the raw data. Finally, the result images can be obtained by the normalization of the transformed wavelet coefficients. Observing the results of their transformations and comparing the images of defects with the actual sizes, this high-precision imaging method could be verified. Therefore, we could make the conclusion that combining THz FMCW and the CWT is a practical high precision imaging method. We believe that the advocated method can promote the development of THz applications.

## 2    Theory

### 2.1 Terahertz FMCW theory

Terahertz FMCW has many common features with the traditional optical FMCW on its basic principle.[9] It is generally accepted that the characteristics of the optical FMCW can be applied to the terahertz band.[10] According to the theory of FMCW, if two beams of terahertz wave are from the same source, they interfere each other with experiencing a different optical path. Therefore,

the intensity of the interference field can be expressed by

$$\begin{aligned} I(\Delta L,t) &= (I_1+I_2)\cdot[1+V\cos(\frac{\alpha\cdot\Delta L}{c}\cdot t+\omega_0\frac{\Delta L}{c})] \\ &= (I_1+I_2)\cdot[1+V\cos(2\pi\cdot f_b t+\varphi_0)] \end{aligned} \tag{1}$$

Where $\Delta L$ is the optical path difference, V is the relative light intensity, $I_1$, $I_2$ are the amplitude,

$\varphi_0$ is the initial phase, $f_b$ is the beat frequency.

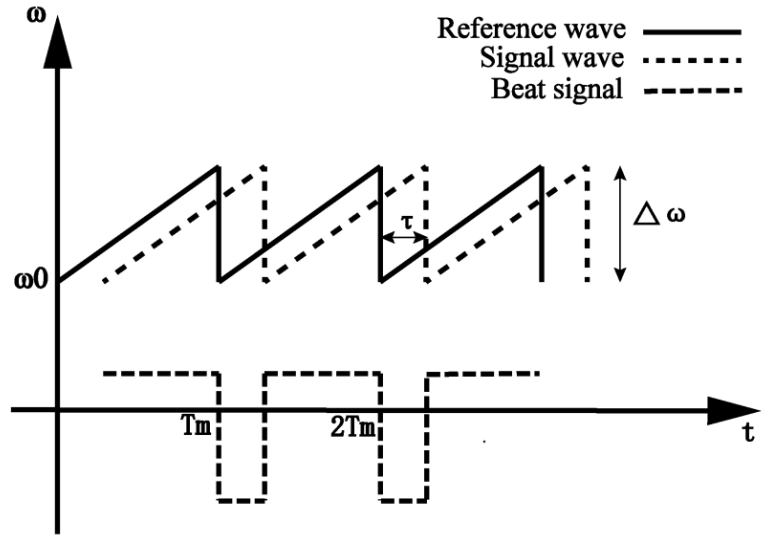

**Fig. 1.** The principle diagram of the THz FMCW. The solid line indicates the reference wave frequency; the dotted
line is the signal light wave frequency; the dotted solid line indicates the beat frequency.

The beat frequency of the interference signal is obtained as:

$$f_b = \frac{\alpha \cdot \Delta L}{2\pi \cdot c} = \frac{\Delta \omega \cdot \Delta L}{2\pi \cdot c \cdot T_m} \quad (2)$$

According to the formula above, we can easily get the relationship $f_b$ and $\Delta L$, furthermore, the $I(\Delta L, t)$ is also can be obtained through Eq.(1). Thus, we get the conclusion that by measuring the beat frequency $f_b$, the position and the reflected light intensity of the target can be calculated [11].

## 2.2 *Continuous wavelet transform algorithm*

In mathematics, a continuous wavelet transform (CWT) is used to divide a continuous-time function into wavelets.[12] Unlike Fourier transform, the continuous wavelet transform has the ability to construct a time-frequency representation of a signal that offers very good time and frequency localization,[13] it is possible for CWT to detect singularities in a signal by using an appropriate basis function $\psi(t)$. Based on the formula of the wavelet, the continuous wavelet transform of the signal $f(t)$ is

$$W_f(a,b) = \frac{1}{\sqrt{a}} \int_{-\infty}^{\infty} f(t)\psi(\frac{t-b}{a})dt \quad (3)$$

The $\psi(t)$ is a continuous basis wavelet function in both the time domain and the frequency domain. The $a$ is the scale factor and the $b$ is translation value.[14]

In this research, the CWT is chosen for signal processing which can provide a better scales selection.[15] The CWT is also continuing to the displacement, during which the analysis wavelet is smoothly moved across the values of the signal. In our experiment, we analyze and process the signals by using a variety of the basis wavelet function $\psi(t)$.

Generally, in order to reduce the computational complexity of signal, the chosen following wavelet functions need to be symmetry and all have simple expressions that are very suitable for the THz signal. Furthermore, they can be continuously differentiated and frequently used in signal identification. Therefore, these three basis wavelet functions are substituted into the equations for the processing of terahertz-reflected signals.

1) Morlet wavelet, the Morlet wavelet is suitable for continuous analysis. There is no scaling function associated with the Morlet wavelet and its mathematical expression can be expressed as

$$\psi(x) = \exp(-\frac{x^2}{2}) \cdot \cos(5x) \tag{4}$$

2) Gaussian wavelet, Gaussian function is a very common function and the derivatives of each order form a wavelet family. Its mathematical expression can be expressed by

$$\psi(x) = \exp(-x^2) \tag{5}$$

3) Mexican hat wavelet, this wavelet is proportional to the second derivative function of the Gaussian probability density function. The wavelet is a special case of a larger family of the derivative of Gaussian wavelets. It can be expressed as

$$\psi(x) = (-1)^2 \frac{d^2 f(x)}{dx^2} = (1 - x^2) \cdot \exp(-\frac{x^2}{2}) \tag{6}$$

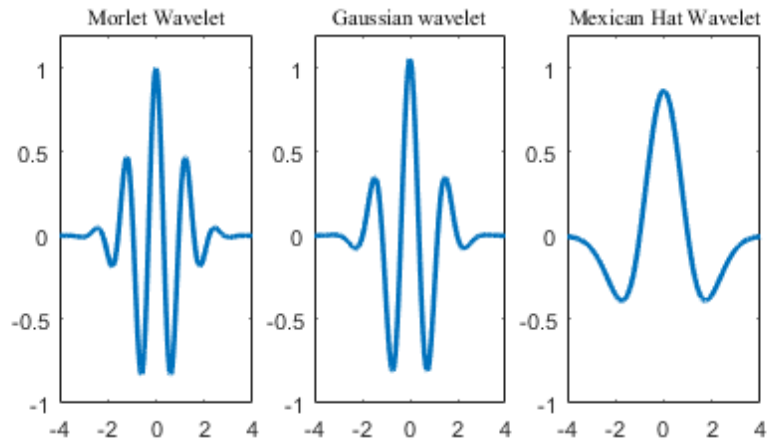

**Fig. 2.** Morlet Gaussian and Mexican hat wavelet waveform. (a) Morlet wavelet waveform; (b) Gaussian wavelet waveform; (c) Mexican hat wavelet waveform.

## 3  Experiment

A terahertz FMCW probe is employed to produce terahertz signals and accept beat frequencies [16-18]. The working principle is shown in Fig 3 (a), where f1, f2 constitutes a focusing lens system, S is the source, ADC is an analog to digital converter, VCO is the voltage controlled oscillator, and D is the heterodyne detector. While it working, the VCO is driven by the ramp generator to produce a signal that sweeps period ($T_m$=240$\mu s$) and the sweep range is approximately 13 to 14 GHz. This signal is multiplied by the light source S, whose sweep range reach 0.23 to 0.32 THz ($\Delta\omega$=90GHz). As for receiving the beat frequencies, a Schottky mixer is used as D. The reflected signal of the object and the delayed transmitted signal are mixed, the mixed-signal is transmitted to the ADC and then is input to the data processing unit. The biggest reason for using this probe is completing the spectral work by the directional coupler, to avoid unnecessary optical devices (reflectors splitters and so on). The output power of this probe is 0.1 milliwatts.[19]

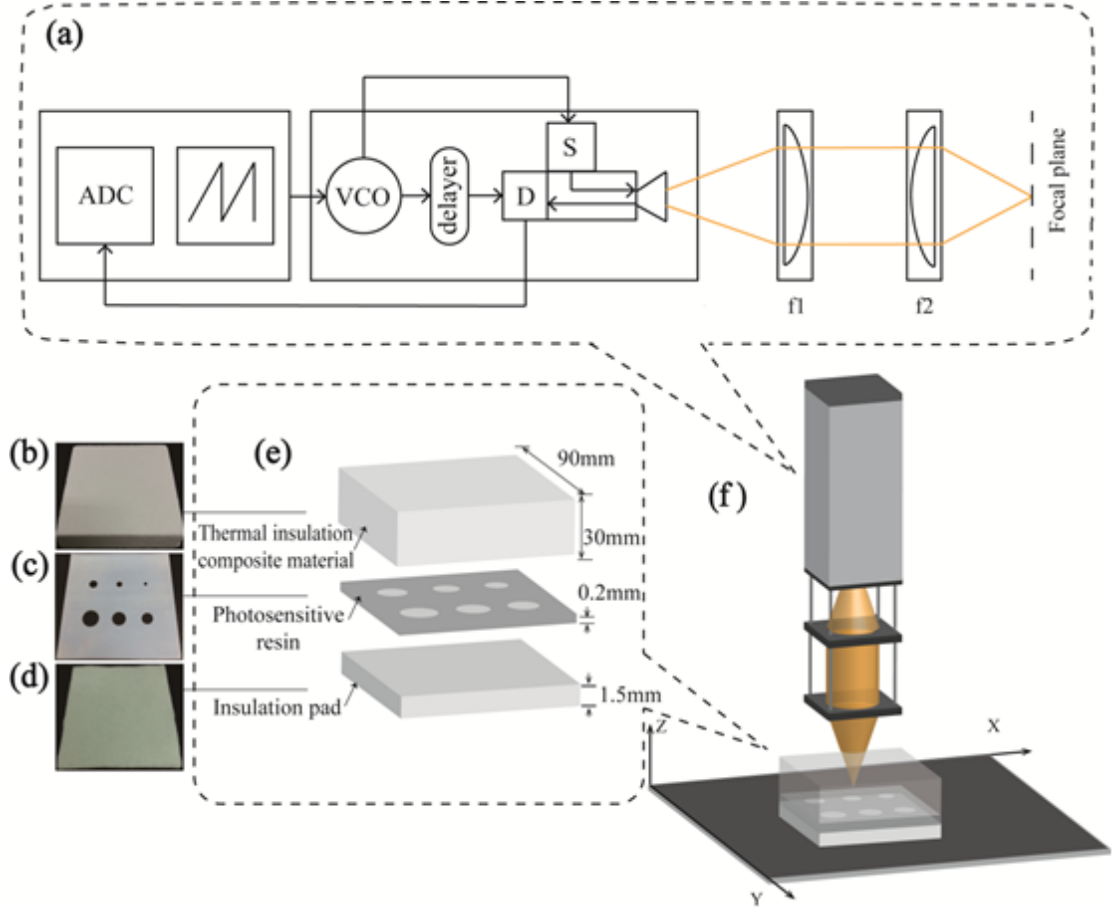

**Fig. 3.** Detection device and samples schematic diagram. (a) Detection device schematic diagram; (b) The image of thermal insulation composite material; (c) The image of photosensitive resin; (d) The image of insulation pad; (e) Schematic diagram of the decomposition of the experimental sample; (f) Schematic of the detection process.

In the experiment, we made a heat shield with six air holes as the experimental sample (air hole defects are most common in heat shield, so we only focus on this type). The heat shield was composed of the thermal insulation composite material, photosensitive resin and insulation pad [shown in Fig. 3(e)]. The photosensitive resin with holes by 3D printed was used to simulate a defective cement layer and its thickness was 0.2mm. The diameter of the holes in the resin was respectively 3mm, 5mm, 7mm, 9mm, 11mm, 13mm. The heat shield of the sandwich structure as shown in the Fig. 3(e) was placed on the reflecting platform, the THz FMCW probe scanned which point by point in XY plane (Because the outgoing light is a Gaussian beam, and its waist diameter is 1mm, the step length is 1mm, to avoid light spot overlap according to the Rayleigh criterion.) Therefore, the in the Z direction of each point was acquired, the experimental data collection is basically completed.[20-21]

## 4    Results and discussion

Using CWT to process THz FMCW signal of composite materials and comparing the effect of three wavelet basis functions is our most important job. In order to verify it clearly, we make a plan includes following steps: Firstly, we pick out two points from good part and defect part to compare the differences of their longitudinal signal. Then we use CWT to process any part to verify its effect on processing the THz signal. Next, we use CWT with three different wavelet basis functions to process defect part and good part to see this method's effectiveness. The final step is using CWT with three different wavelet basis functions to process all points and comparing their slice pictures, the quantitative results are listed in a table. We will get the conclusion by analyzing the final data.

Before analyzing, we make a simple visualization of the slice display shown in Fig. 4. According to the content shown by the slice image, we select the two feature points A and B along the Z axis to compare the reflection difference of longitudinal signal between the defect part B (air holes) and the good part A (photosensitive resin), which are plotted as curves in Fig. 5(a)(b).

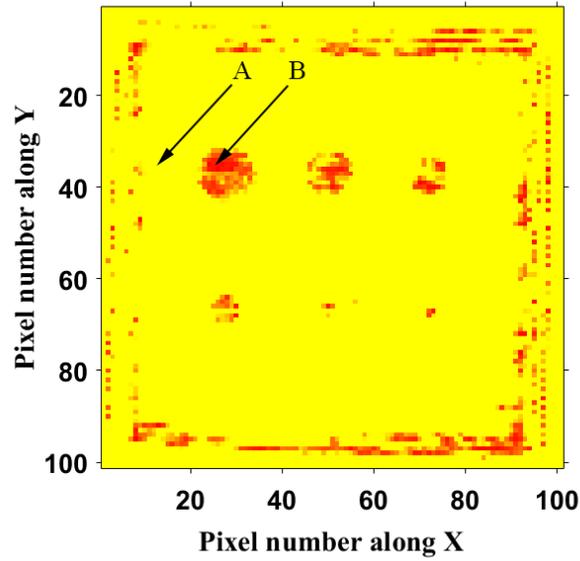

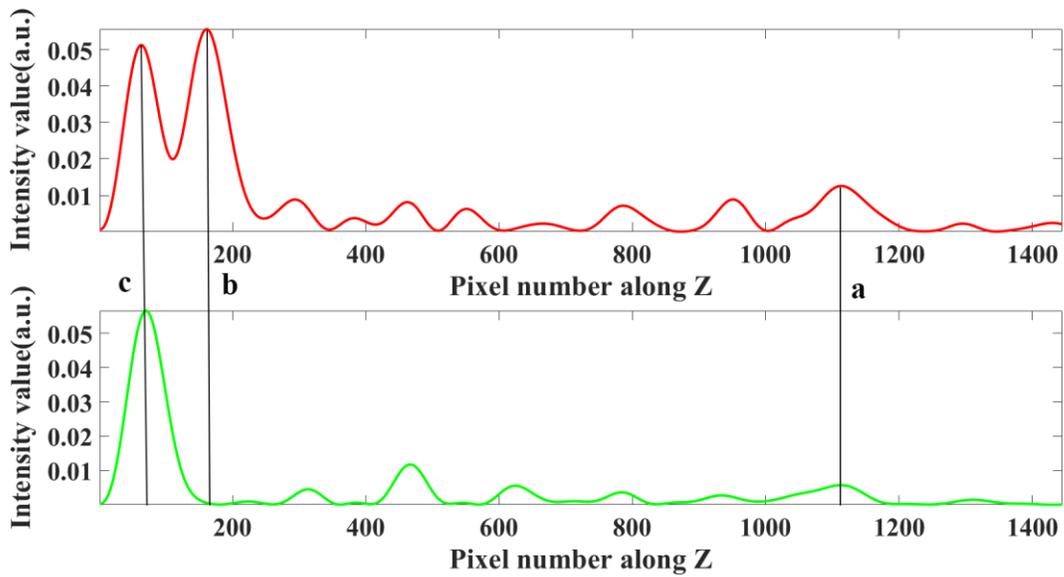

**Fig. 5.** The reflection curves of characteristic points. (a) The reflection curves of A(upper); (b) The reflection curves of B(under).

Through the observation of Fig. 5(a)(b), we could figure out the obvious information in the curve. The location of dashed line 'a' represents the interface between the thermal insulation composite material and the air. The position of dashed line 'b' indicates the interface between the lower surface of the thermal insulation composite material and air holes or the photosensitive resin. The peak represents relatively stronger reflection feature, which is the difference between 'A' and 'B' at dashed line 'b'. The location of dashed line 'c' indicates the reflecting platform, A and B curve could be regarded the same in this position. There are some weak peaks between the positions of the dashed lines 'a' and 'b' because of the unevenness of the interior of heat shield or other factors, and there are also some overlaps of the waveform, which may interfere with our identification of the signal at the interface. However, the CWT has a very good ability to distinguish the weak signals from the noise and separate the overlapping waveform. Using CWT could improve the accuracy of our measurements of the waveform at each point.

In order to choose an appropriate scale of the wavelet to analyze this terahertz signal, the width of the wavelet should be limited to the peak of the signal. According to Fig. 6, the peaks of the signal that we concerned is not only strengthened but also separated due to the effect of the CWT, and overlapping waveform almost disappears. Therefore, the information of the defects could be clearly identified and the layer at the interface that we focus on is also easily picked out. Because of the CWT to optimize the waveform of each point, the contrast of the XY slice image has also been improved.

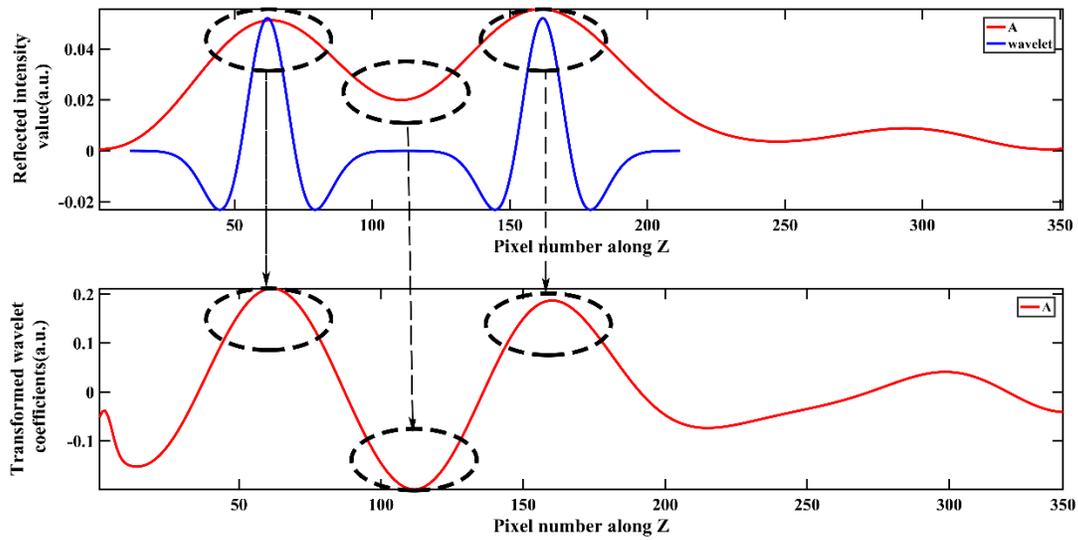

**Fig. 6.** The schematic diagram of the reflected signal is optimized and strengthened by wavelet transform. (a) The schematic diagram of the reflected signal is strengthened by wavelet transform(upper); (b) The image after the Mexican Hat wavelet transforming(under).

We still take the waveform of A and B as the example to carry out different basis functions. As shown in Fig. 7, three different wavelet transforms have similar effects on A and B signals. Obviously, after the transformation, the interface is distinguished and the information of the details have been strengthened.

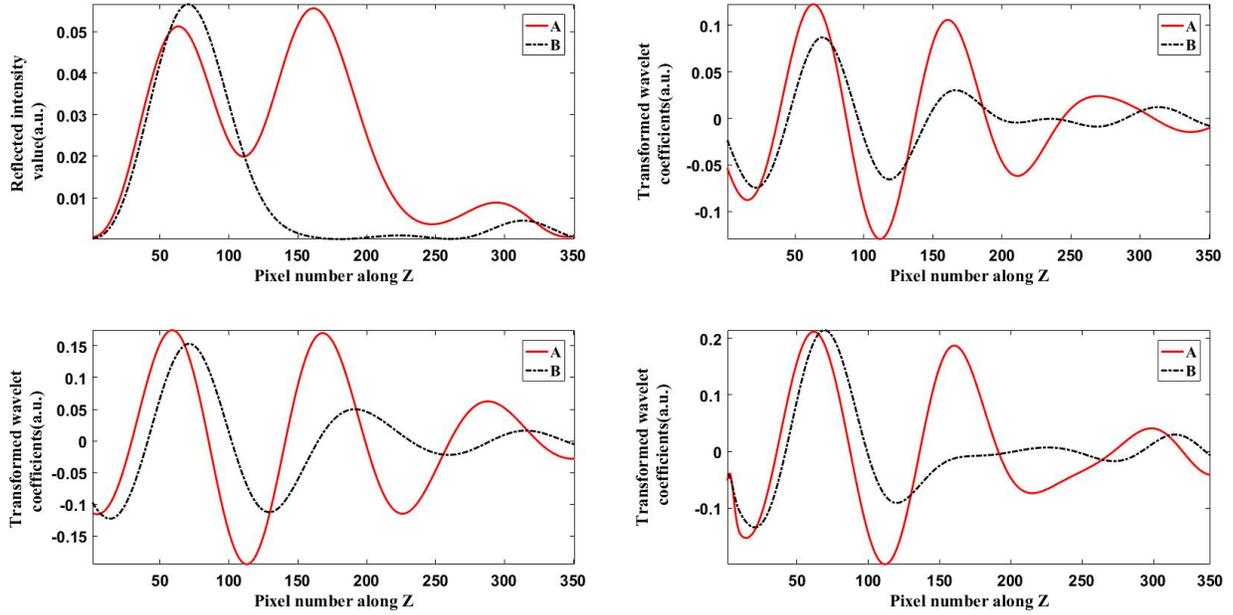

**Fig. 7.** The original waveforms of A and B points and the wavelet coefficient curves of three different basic functions. (a) The original waveforms of A and B points; (b) The curves of A and B points after Morlet wavelet transformed; (c) The curves of A and B points after Gaussian wavelet transformed; (d) The curves of A and B points after Mexican Hat wavelet transformed.

We use the CWT algorithm with three basic functions for all point to construct the three-dimensional volume data sets, therefore, the sliced image at the simulated cemented layer is correspondingly enhanced. In the experiment, the transformed wavelet coefficients normalized to one as the original data, and then we select the layer where the air holes and thermal insulation composite material intersect (position 'b') to draw the images.

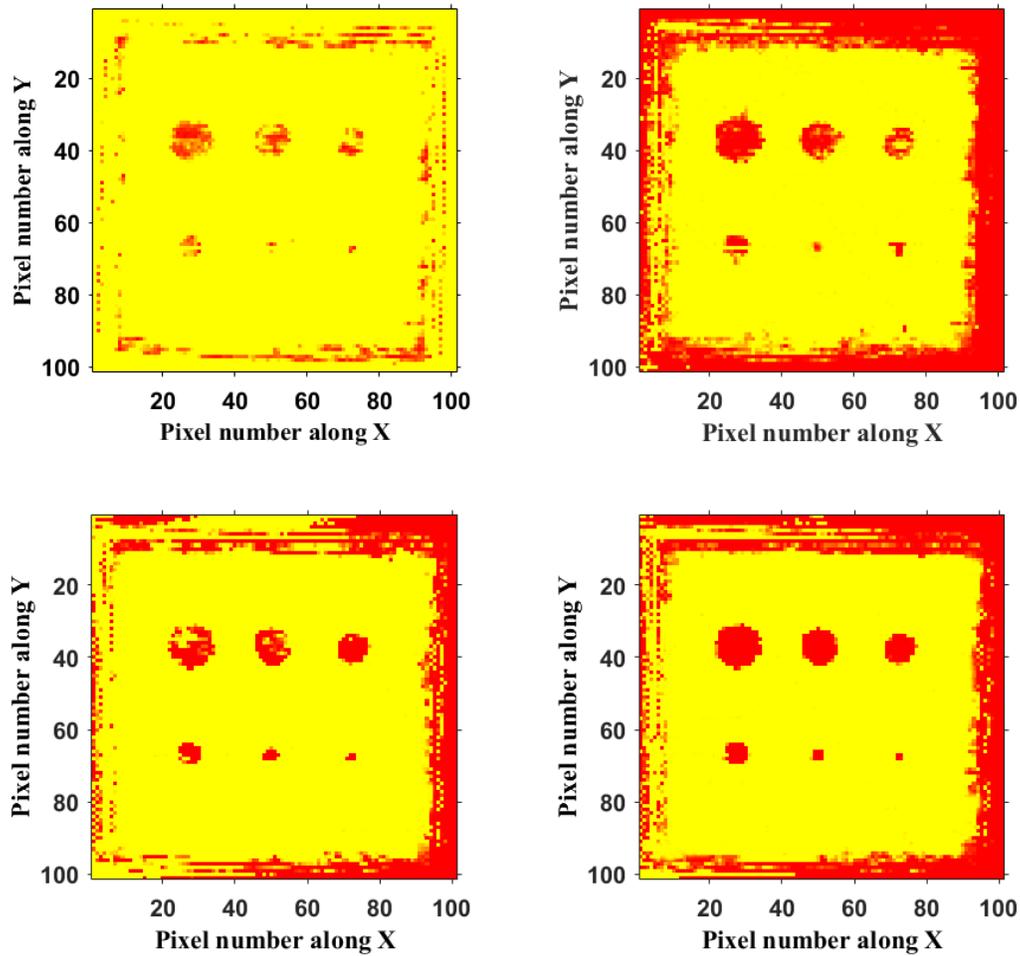

**Fig. 8.** The original slice picture and the slice picture after three different transformed. (a) The original slice picture; (b) The slice picture after Morlet wavelet transformed; (c) The slice picture after Gaussian wavelet transformed; (d) The slice picture after Mexican Hat wavelet transformed.

As can be seen from Fig. 8, after the CWT, the defects in the image are obviously more clear than the original image and their edge and details are more prominent. We then use some image processing algorithms to calculate the area of defects in each slice image and compare it with the original image to show that the continuous wavelet transform does improve the precision of imaging. The calculation results are shown in Table. 1.

**Table 1. Calculation results of defects in the experiment**

| Area / Slice images | defect 1 ($mm^2$) | defect 2 ($mm^2$) | defect 3 ($mm^2$) | defect 4 ($mm^2$) | defect 5 ($mm^2$) | defect 6 ($mm^2$) | Total difference ($mm^2$) | Percent difference |
|---|---|---|---|---|---|---|---|---|
| Actual size | 7.065 | 19.625 | 38.465 | 63.585 | 94.985 | 132.665 | 0 | 0% |
| Raw data | 6 | 8 | 34 | 53 | 85 | 112 | 58.39 | 16.38% |
| Morlet | 6 | 10 | 42 | 61 | 95 | 133 | 17.16 | 4.81% |
| Gaussian | 8 | 20 | 35 | 59 | 90 | 120 | 27.01 | 7.58% |
| Mexican hat | 7 | 17 | 36 | 63 | 96 | 133 | 7.09 | 1.99% |

According to Table 1, the results show that the data processed by the CWT algorithm can effectively enhance the contrast of THz imaging. By comparing the transformed measurement results with the raw data, we find that results calculated by Mexican hat wavelet transformed are closest to the actual size, whose total difference of the six defects is 7.09 and only account for 1.99% of the total area. The other two wavelet functions also improve precision, the total difference of Gaussian function is 27.01 and the area difference rate is 7.58%. The total difference by Morlet function is 17.16 and difference rate is 4.81%.

## 5    Conclusion

In our study, FMCW theory is applied to the THz band to detect the defects inside the heat shield. The experiment confirms that it is effective for THz nondestructive testing of insulating composites, which not only improves the accuracy but also time efficient. Then, the raw data collected by FMCW system is processed by CWT with three wavelet basis functions, after that, the quality of images has been a certain improved. Comparing the results listed in the table, we find that three wavelet basis functions in our paper all improve the accuracy of THz imaging, the area difference rate of which respectively reach to 4.81%, 7.58%, 1.99%. Among them, the Mexican hat wavelet function performs best in processing THz signals. And because of the difference in expressions, their ability to distinguish the defects of different sizes may be different, its worth further study in the future work. Therefore, we can summarize that it is a new effective way to improve the accuracy of THz imaging that by combing the FMCW theory and the CWT, and the accuracy can be further improved by changing the wavelet basis function. We believe our method can have widespread applications in terahertz imaging, not just for the detection of aerospace composite materials, but also in other areas such as security check and archaeology detection.

*Acknowledgments*

Funding Information. We gratefully acknowledge the support of the National Natural Science Foundation of China (No. 61405063, 61475054, 11574105，61177095), and the Hubei Science and Technology Agency Project (No. 2015BCE052).

*Author Biographies*

**Yu Zhou** is a master student at the Huazhong University of Science and Technology. His current research interests include terahertz imaging, image processing, and optoelectronic systems.

*Tables*

**Table 1. Calculation results of defects in the experiment**

| Area Slice images | defect 1 ($mm^2$) | defect 2 ($mm^2$) | defect 3 ($mm^2$) | defect 4 ($mm^2$) | defect 5 ($mm^2$) | defect 6 ($mm^2$) | Total difference ($mm^2$) | Percent difference |
|---|---|---|---|---|---|---|---|---|
| Actual size | 7.065 | 19.625 | 38.465 | 63.585 | 94.985 | 132.665 | 0 | 0% |
| Raw data | 6 | 8 | 34 | 53 | 85 | 112 | 58.39 | 16.38% |
| Morlet | 6 | 10 | 42 | 61 | 95 | 133 | 17.16 | 4.81% |
| Gaussian | 8 | 20 | 35 | 59 | 90 | 120 | 27.01 | 7.58% |
| Mexican hat | 7 | 17 | 36 | 63 | 96 | 133 | 7.09 | 1.99% |

Fig. 1. The principle diagram of the THz FMCW. The solid line indicates the reference wave frequency; the dotted line is the signal light wave frequency; the dotted solid line indicates the beat frequency.

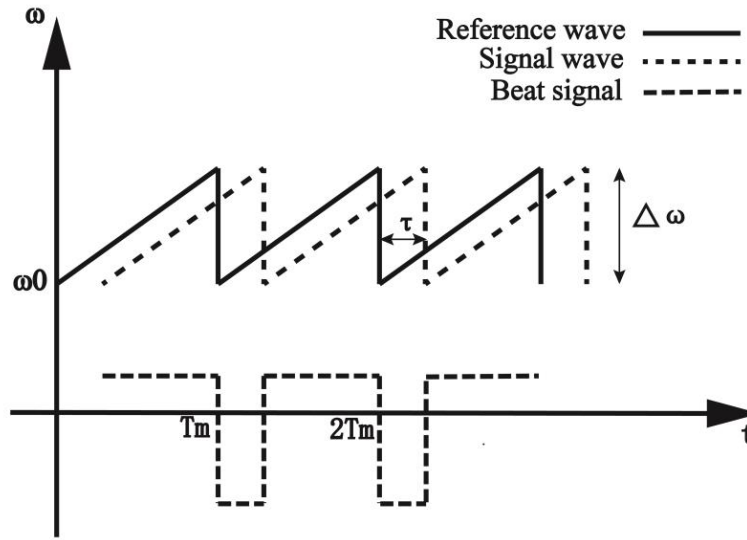

Fig. 2. Morlet Gaussian and Mexican hat wavelet waveform. (a) Morlet wavelet waveform; (b) Gaussian wavelet waveform; (c) Mexican hat wavelet waveform.

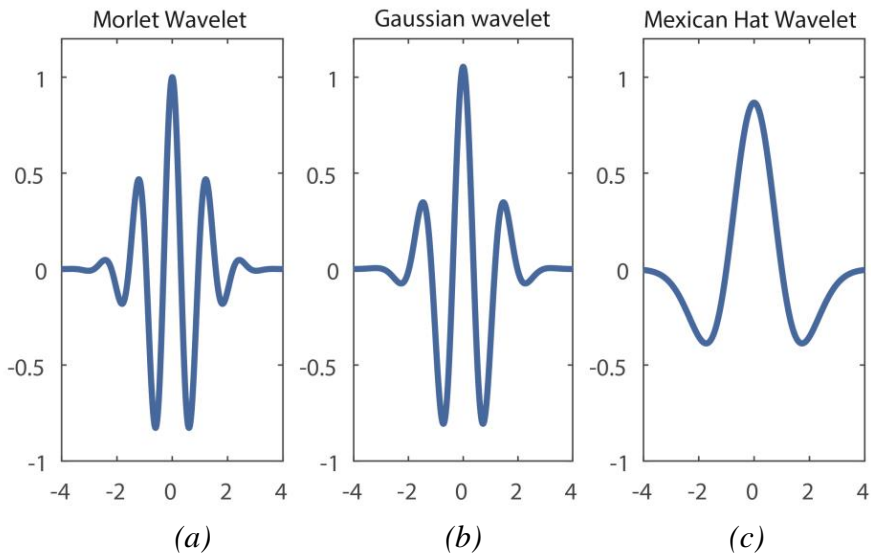

Fig. 3. Detection device and samples schematic diagram. (a) Detection device schematic diagram; (b) The image of thermal insulation composite material; (c) The image of photosensitive resin; (d) The image of insulation pad; (e) Schematic diagram of the decomposition of the experimental sample; (f) Schematic of the detection process.

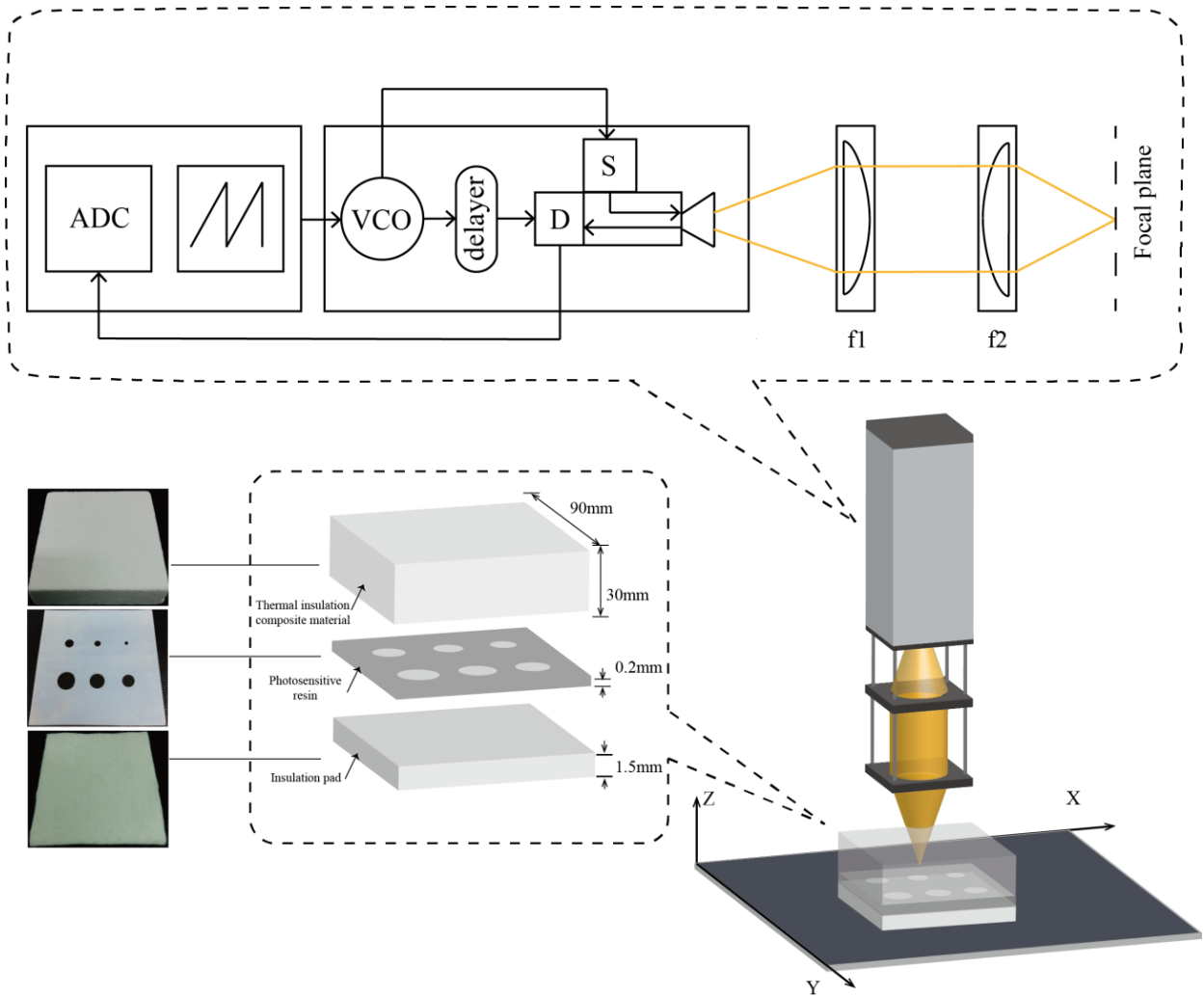

Fig. 4. Slice image of raw data.

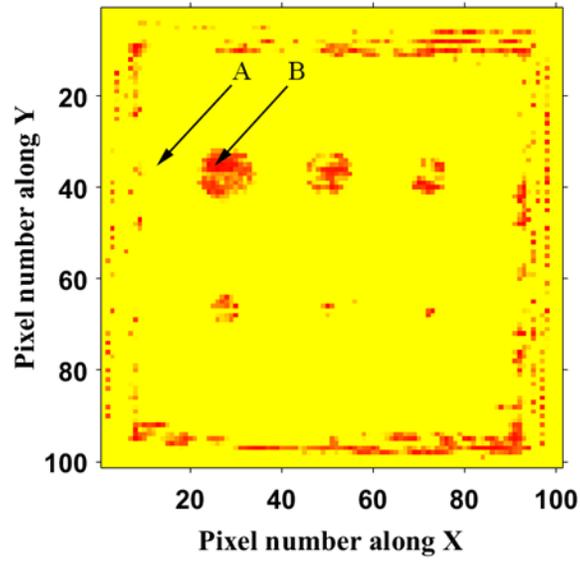

Fig. 5. The reflection curves of characteristic points. (a) The reflection curves of A(upper); (b) The reflection curves of B(under).

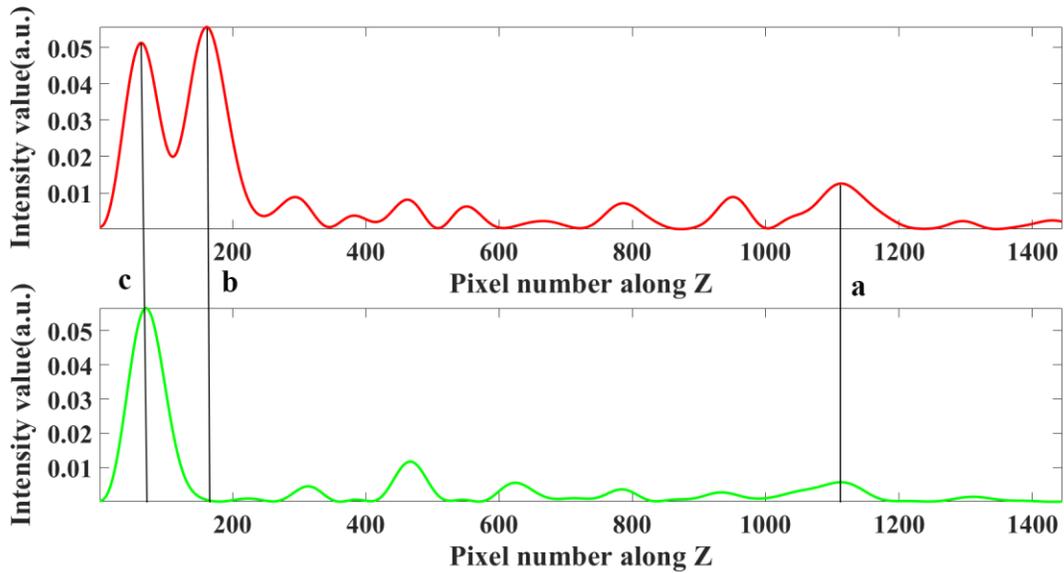

Fig. 6. The schematic diagram of the reflected signal is optimized and strengthened by wavelet transform. (a) The schematic diagram of the reflected signal is strengthened by wavelet transform(upper); (b) The image after the Mexican Hat wavelet transforming(under).

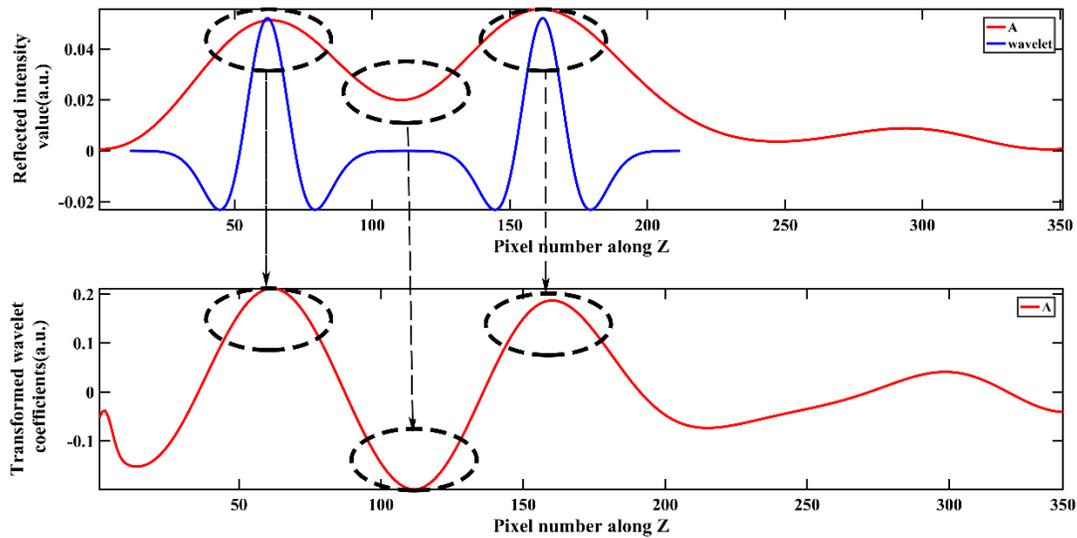

Fig. 7. The original waveforms of A and B points and the wavelet coefficient curves of three different basic functions. (a) The original waveforms of A and B points; (b) The curves of A and B points after Morlet wavelet transformed; (c) The curves of A and B points after Gaussian wavelet transformed; (d) The curves of A and B points after Mexican Hat wavelet transformed.

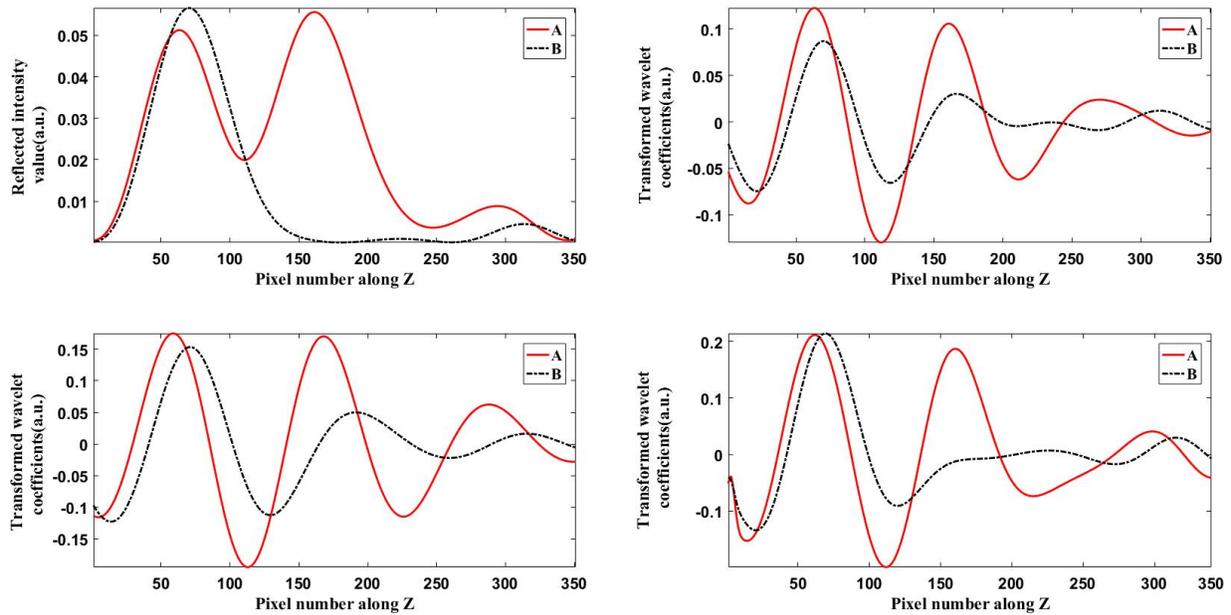

Fig. 8. The original slice picture and the slice picture after three different transformed. (a) The original slice picture; (b) The slice picture after Morlet wavelet transformed; (c) The slice picture after Gaussian wavelet transformed; (d) The slice picture after Mexican Hat wavelet transformed.

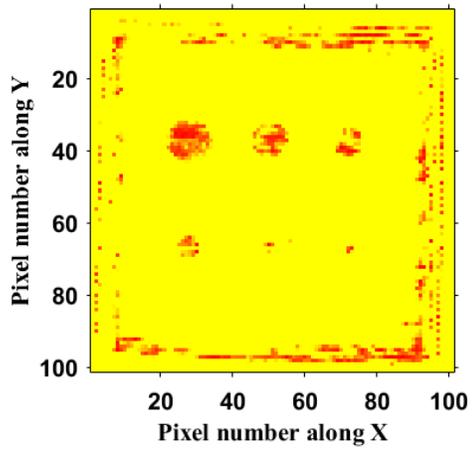

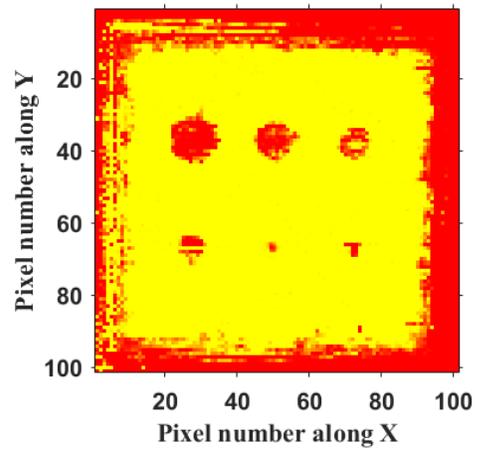

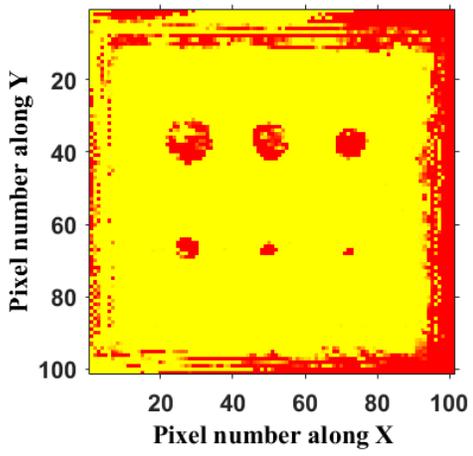

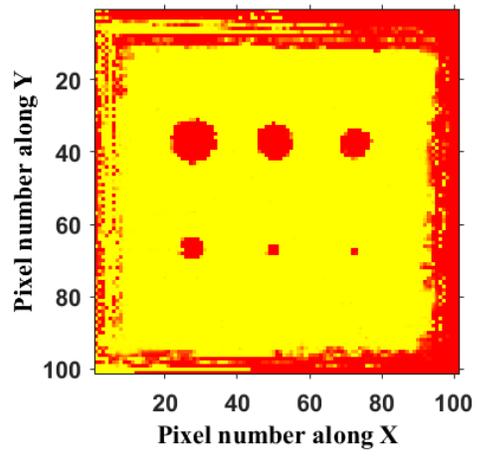